# Limitations due to residual interference in a fiber-based optical frequency reference at 1.55 µm

K. Manamanni,1 T. Steshchenko,1 F. Wiotte,1 R. Le Targat,2 M. Abgrall,2 O. Lopez, 1 E. Cantin,1 P-E. Pottie,2 A-. Amy-Klein1, V. Roncin,1 F. Du-Burck,1, *

*1-Laboratoire de Physique des Lasers, CNRS, Université Sorbonne Paris Nord, 99 av. J.-B. Clément, F-93430 Villetaneuse, France*

*2-LNE-SYRTE, Observatoire de Paris, Université PSL, CNRS, Sorbonne Université, 61 Avenue de l'Observatoire, F-75014, Paris, France*

* Corresponding author: frederic.du-burck@univ-paris13.fr

**Abstract**

We present an experimental investigation of the stability limits specific to optical frequency standards using a fiber-optic architecture and semiconductor lasers. A compact setup composed of a semiconductor laser frequency-locked onto an acetylene transition detected in saturated absorption has been implemented using only fiber-optic components. Fiber optic technology allows compact and reliable solutions for various applications. However, for high sensitivity and stability applications such as metrology, residual reflections induced by optical index inhomogeneities in connectors and fiber-coupled components leading to spurious interference significantly limit performance. We have examined the origin of the interference fringes superimposed on the detected signal and the limitations they cause to the frequency stability of the reference. The effects of temperature and beam power fluctuations are also examined. Our results show that the frequency stability is limited in the $10^{-13}$ range by the effect of interference fringes due to use of fiber components.

## 1- Introduction

Only a few molecules present transitions in the extended fiber-optic telecommunications range. Several frequency references realizations have been reported from 1260 to 1675 nm using $C_2H_2$, HCN, CO, $CO_2$, $H_2S$, $CH_4$, $H_2O$ [1]. The most popular molecule is acetylene owing to the (ν1 + ν3) overtone band of acetylene $^{13}C_2H_2$ within the C band (1530–1560 nm). Moreover, it occupies a special place for its P(16) transition at 1542.38 nm, which has been chosen by the CIPM as a primary wavelength standard [2].

Thus, acetylene spectroscopy in this region has been carefully studied and high stability and accuracy references have been demonstrated, based on the detection of sub-Doppler resonances in acetylene vapor probed in a cell [3, 4, 5] or in a cavity [6, 7, 8; 9]. In both configurations, frequency stabilities (the relative frequency fluctuations) in the $10^{-14}$ range have been reached [5, 9].

In parallel to these laboratory experiments, special attention is given to the development of compact and portable references dedicated to non-laboratory measurements. For this purpose, 1550 nm fiber-optic components allow considerable reduction of the volume of such acetylene-based frequency standards. In addition to the compactness, this approach guarantees the robustness of such setups for optical alignments. However, several issues inherent to the use of optical fibers and photonics components limit their stability performance. Residual Fresnel reflection in connectors and diopters leading to interference fringes, as well as the sensitivity of light polarization with the mechanical and thermal environment are the main issues reported in the literature.

For example, Ahtee et al. [10] describe the performance of an acetylene-stabilized laser system based on a fiber setup, in which the acetylene cell is the only part of the setup in free space. To overcome the problem of interference, the optical path length is dithered using a vibrating mirror in the free space part of the setup. A fractional frequency deviation of $8.8\ 10^{-12}\ \tau$-1/2 for integration time in the range 10 - 1000 s is reported.

Initiated by the development of hollow-core photonic crystal fibers (HC-PCF), more compact acetylene cells could be realized by filling such fiber with gas and hermetically connecting it at both ends to standard single-mode fibers [11, 12]. This makes a promising all-fiber setup with intermediate stability performance limited principally by HC-PCF guiding issues and air/silica reflections in the cell [13, 14].

In this work, we present the experimental setup of a fiber-based optical frequency reference consisting of a semiconductor laser locked onto the P(16) transition of the (ν1 + ν3) overtone band of $^{13}$C2H2 at 1542.38 nm detected in a cell. The saturated absorption signal is detected by a modulation transfer technique in which the saturating beam is frequency modulated and the detection is performed on the probe beam at the modulation frequency. This results in a background-free sub-Doppler signal. A stability in the $10^{-13}$ range between 1 and $10^4$ s is demonstrated. Various effects limiting the stability are analyzed and it is shown that the essential limitation is due to interference whose origins are carefully identified.

## 2- Experimental set-up

The principle of the experiment is depicted in Fig. 1. The part of the setup limited by the dotted line corresponds to the free space part between two collimators. The laser diode is a Rio Planex with a linewidth of 10 kHz and an output power of 12 mW for a 120 mA current. Its output power is split into two beams, one to saturate the absorption of the acetylene vapor in the cell and the other propagating in the opposite direction in the cell for probing the absorption.

The saturating beam is frequency shifted ($f_1$ = 25 MHz) and frequency modulated at $f_{FM}$ = 200 kHz with a modulation index $\beta$ = 2 by an acousto-optic modulator (AOM#1). The saturating beam is amplified by an Erbium-Doped Fiber Amplifier (EDFA) and its power in the cell is automatically controlled by a variable optical attenuator driven electronically (VOA#1). This setup compensates for power variations induced by polarization fluctuations upstream of a polarization beam splitter (PBS#1) which ensures a linear polarization in the cell. We have verified that the power variations detected on Photodiode #2 were not due to interference after PBS#1. The power of the saturating beam is 30 mW and the corresponding saturation parameter is ~0.4.

The probe beam is frequency shifted ($f_2$ = 40 MHz) by the acousto-optic modulator AOM#2 in order to avoid the detection of interference due to the residual zero-order at AOM#1's output as will be discussed later. The probe power is automatically controlled in the same way as the saturating beam by the electronically driven variable optical attenuator VOA#2. The polarization fluctuations upstream of PBS#2 are then compensated. An optical circulator is used to extract the saturating beam at the cell's output. The probe power in the cell is 2 mW. The probe power modulated by the absorption in the cell at $f_{FM}$ = 200 kHz is transmitted into PBS#1 and measured by Photodiode#1. The probe and pump polarizations are maintained orthogonal thanks to the polarization controller PC#2 upstream of PBS#2.

The free-space part of the setup is composed of two adjustable collimators (Schäfter + Kirchhoff), the cell filled with acetylene ($^{13}C2H2$) with a pressure of 2.5 Pa, and two mirrors for beam alignment (not shown in Fig. 1). The radius of both beams is about 0.4 mm for a fiber-to-fiber optical power coupling rate of 70 %. The cell is made and filled by the Institute of Scientific Instrument (Brno, Czech Republic). Its diameter is 3 cm and its length is 30 cm. An anti-reflection coating is deposited on each optical window of the cell.

The setup is into a box made of high-density polyurethane plates ensuring basic thermal and acoustic insulation, named hereafter "thermal box". The temperature in the lab is controlled at 19 ± 1° C by an air conditioning system.

The error signal is detected at fFM = 200 kHz by a lock-in amplifier. The detected signal is applied to a proportional-integral controller (PI controller) in order to generate the laser frequency correction signal. Two integrators are cascaded. The first one generates a fast correction signal acting on the laser current in a correction band extending up to 2 kHz. The second integrator corrects for slow fluctuations by acting on the set-point of the laser temperature control.

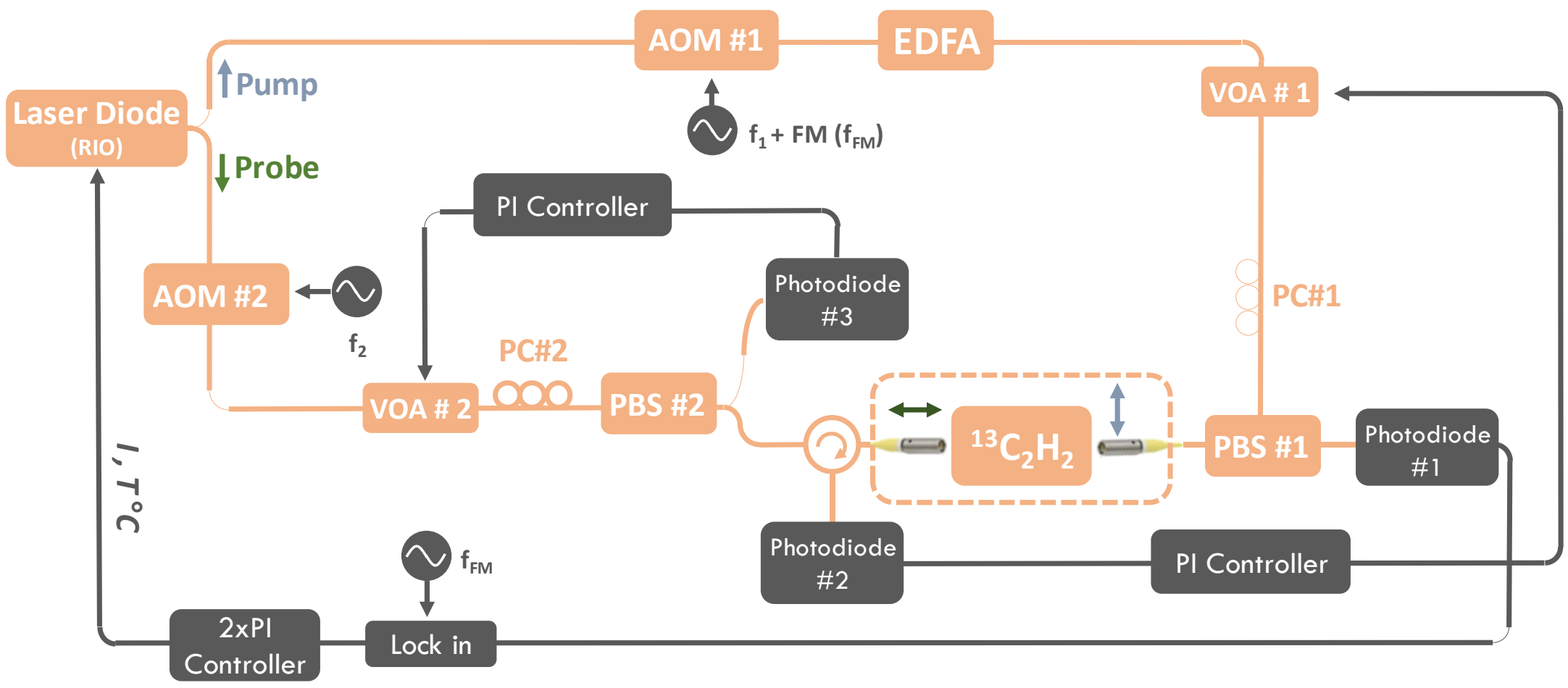

Fig. 1. Laser stabilization setup (AOM: Acousto-optic modulator; PBS: Polarization beam splitter; VOA: Variable optical attenuators; PI controller: Proportional-integral controller; PC: Polarization controller).

The spectral analysis of the demodulated signal at 200 kHz, used for frequency stabilization of the laser on the acetylene transition at 1542.38 nm is shown in Fig. 2. The green curve is the detection noise that fixes the sensitivity of the measurement. The black curve is obtained by reducing the servo bandwidth (the frequency range in which the correction is active) down to 200 Hz. The trace thus corresponds to the open-loop frequency noise for Fourier frequencies higher than 200 Hz from which it reaches the $1/f^2$ trend. The red curve is the noise in closed loop. We also plot in Fig. 2 the probe beam intensity noise measured by shifting the laser frequency out of the transition. This intensity noise is converted into frequency noise (blue curve) from the slope of the detected signal (inset of Fig. 2).

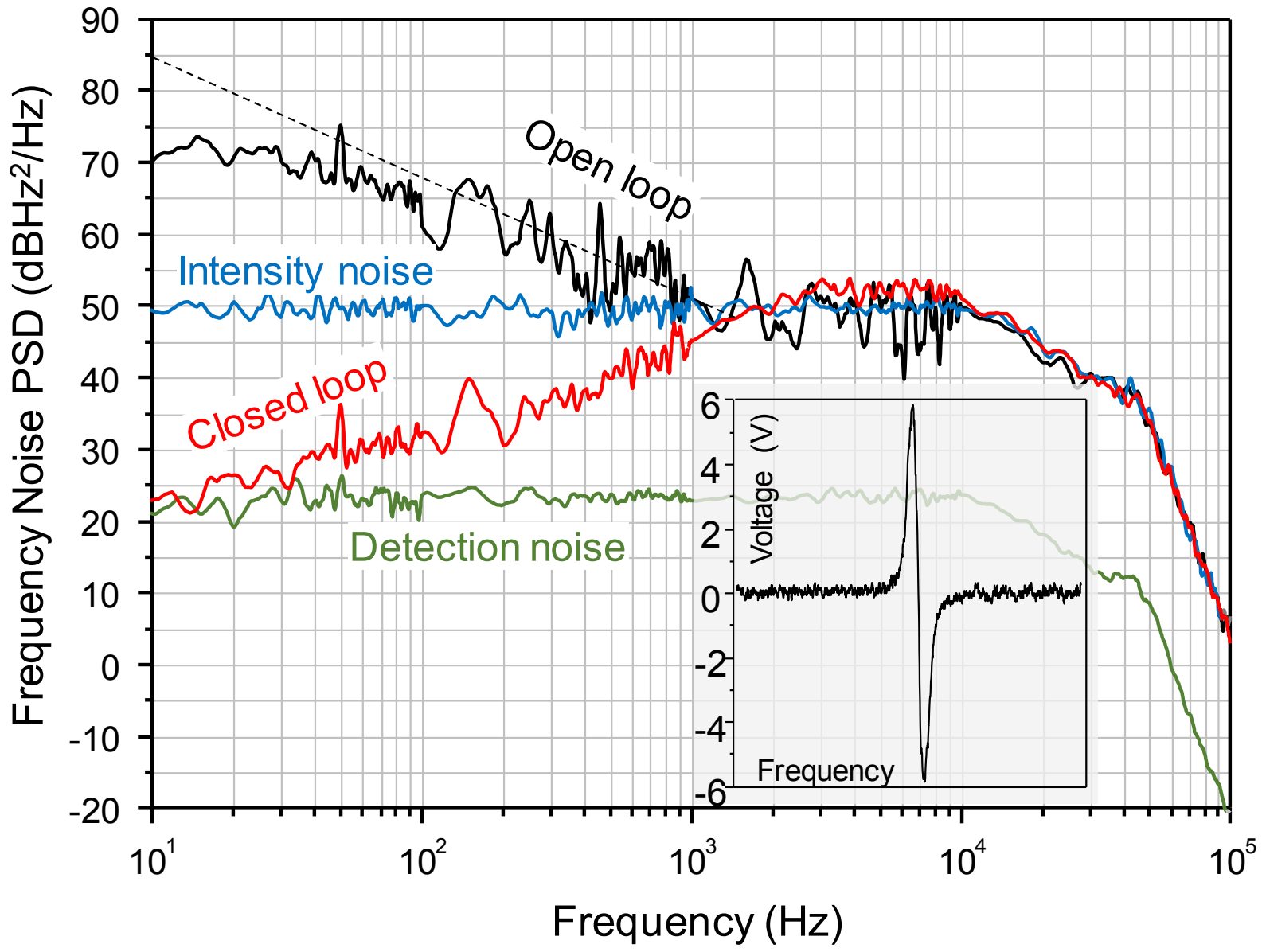

Fig. 2. Power spectral density of the error signal for frequency stabilization of the laser onto the acetylene transition at 1542.38 nm. The black curve is the open-loop noise (for frequency > 200 Hz – see text), the red curve is the closed loop noise, the blue curve is the intensity noise contribution limiting the noise rejection, the green curve is the detection noise giving the sensitivity of noise measurements. Inset: trace of the error signal corresponding to the detected molecular profile.

The noise reduction above 50 kHz observed in all these curves is due to the mixer output filter. The measured closed-loop noise reaches the open-loop noise curve at the frequency 2 kHz, which therefore corresponds to the servo bandwidth. The probe beam intensity noise is white noise at the 50 $dBHz^2/Hz$ level. Throughout the correction band, this noise level is higher than the noise measured in closed loop. Consequently, the actual frequency noise of the corrected laser copies this intensity noise, which limits the possibilities of correction.

The error signal is shown in the inset of Fig. 2. Its peak-to-peak width is 630 kHz and its amplitude is 12 V leading to a frequency-to-voltage conversion factor of ~2 $10^{-5}$ V/Hz. The sensitivity of the laser frequency to the input voltage is ~ 37.5 MHz/V and the cut-off frequency of the PI corrector is 1.6 Hz. This results in a servo bandwidth of ~1.2 kHz in good agreement with the experimental value obtained in Fig. 2.

**3- Stability results**

The stability of our frequency standard at 1542.38 nm is evaluated from a comparison with a metrological reference provided by LNE-SYRTE (Observatoire de Paris) in the framework of the national metrological network REFIMEVE+ [15]. A continuous wave Erbium-doped fiber-laser emitting at $\lambda$ref = 1542.14 nm (channel 44 of WDM ITU-T) is stabilized onto an ultra-stable Fabry-Perot cavity [16]. For long-term stability, it is referenced onto a hydrogen-maser, ensuring a stability better than $10^{-15}$ until 105 s of integration time. The laser is transferred to our lab through a 43 km long single mode fiber. The propagation noise along the fiber, mainly thermal and acoustic noise, is actively compensated [17]. In our lab, the signal is regenerated by phase locking a local laser to the incoming reference signal and is then distributed to our experiment using a multibranch laser station [18]. As a result, the transfer does not degrade the stability and accuracy of the transmitted optical reference [17,18].

The frequency shift between the metrological reference and the C2H2 stabilized laser is about 30 GHz. For stability measurements, both sources are modulated in the same Mach-Zehnder modulator at fs = 15 GHz, generating two sets of sidebands. By tuning the modulation frequency fs, we obtain a beat-note at 60 MHz that is filtered and sent to a frequency counter. In order to quantify the frequency stability of the reference, Allan deviations in relative value $\sigma_y$ (fractional Allan deviation) of the beat-note are drawn in Fig. 3.

The three curves in Fig. 3 reflect the impact of temperature and power fluctuations on the frequency stability of the setup described in Fig. 1. For integration times less than 10 s, all three curves exhibit the same fractional stability of 8 $10^{-13}$ $\tau^{-1/2}$, consistent with the limita on of fre uency correc on due to the laser white amplitude noise iden ed in ig. . Indeed, this noise of 50 dBHz$^2$/Hz leads to an Allan deviation of the form $10^{-12}$ $\tau$ -1/2 [19], in good agreement with the experimental stability obtained in Fig. 3.

For integration times higher than 10 s, the evolutions of the stability curves are different, but some bumps are observed on the three curves. We will show in the next section that this behavior is due to interference fringes coming from parasitic reflections on fiber components.

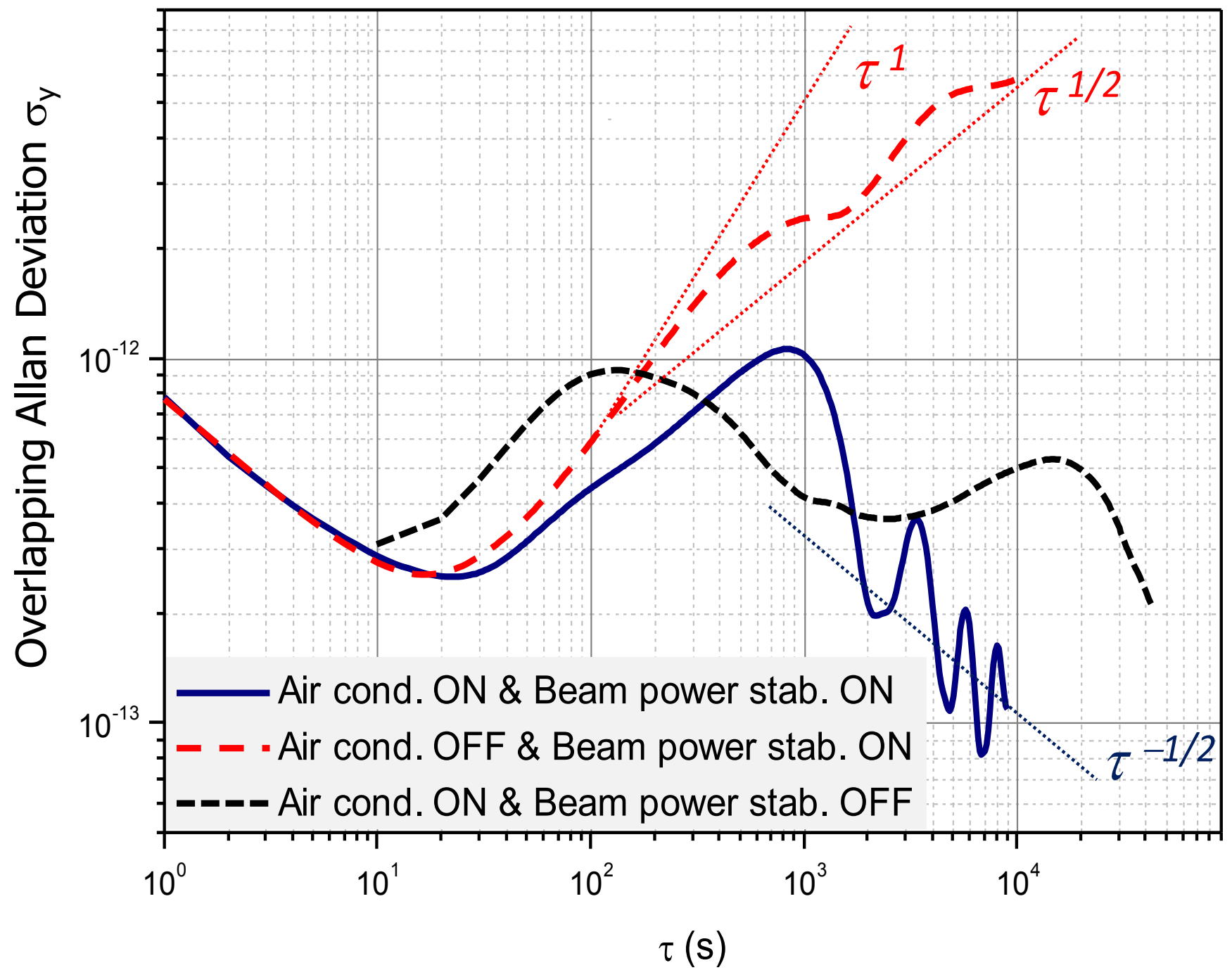

Fig. 3. Stability characterization of the fiber-based frequency standard at 1.55 µm with fractional Allan deviation (beat-note with the optical reference transferred through the REFIMEVE+ network). Three measurements are achieved, two concerning the thermal sensitivity thanks to the air conditioning system in the lab (blue and red curves) and one concerning the power sensitivity thanks to power beam stabilization system using VOAs (black curve).

The blue solid line curve is obtained for a temperature in the lab controlled by the air conditioning system and when the power of both beams are automatically controlled by VOA#1 and VOA# in such a way that it is stable at the two PBS's outputs. or integration times greater than 10 s, a characteristic shape resulting from periodic variations of the laser frequency with a period slightly greater than 2000 s is clearly visible. Note that the level of stability always remains in the $10^{-13}$ range and that the bumps due to periodic frequency fluctuations are superimposed on a general decreasing trend that reaches a stability level in the $10^{-14}$ range at $10^4$ s.

The origin of these periodic stability variations, analyzed in Section 4, is attributed to interference fringes due to parasitic reflections on the fibered components of the setup.

The red curve in long dashes is obtained when the air conditioning system is turned off, which generates temperature fluctuations during the long-term stability measurements. We will show in Section 5 hereafter that the temperature sensitivity of our setup is clearly caused by the thermal sensitivity of AOM#1.

The black curve in short dashes shows the sensitivity of the setup to beam power stability. It is obtained when the air conditioning system is on but without power stabilization at the output of PBS's. The sensitivity to beam power is analyzed in Section 6.

**4- Interference analysis**

The most significant limitation observed in fiber-based frequency references comes from beam reflections in the setup, at the interfaces of the fiber components and on the collimators [10]. These various reflections result in interference fringes added to the output signal of the photodiode monitoring the probe beam. However, only those modulated at $f_{FM}$ = 200 kHz are detected by the lock-in amplifier and superimposed to the error signal. These interference fringes lead to fluctuations in the zero-crossing of the detected signal and thus, to an instability of the optical frequency of the standard.

When the acousto-optic modulator AOM#2 shifting the probe beam frequency by f2 = 40 MHz is removed, significant interference fringes representing 8% of the amplitude of the error signal appear (Fig. 4(a)). They are necessarily due to the intense saturating beam, the only one to be modulated in the modulation transfer method. However, we observed that these interference fringes exist only when the probe beam is also received by the photodiode. Consequently, these interference fringes do not result from the direct detection of the residual reflections of the saturating beam, which would then exist independently of the probe beam. They result from the homodyne mixing on the photodiode of the probe beam and the residual zero-order output beam of AOM#1, whose intensity is measured at 50 dB below that of the first-order output beam. This process is illustrated in Fig. 5(a). Of course, the zero-order beam cannot be frequency-modulated. However, it is observed that the first-order output beam carries residual amplitude modulation (RAM) at $f_{FM}$ at the level of 2 $10^{-3}$ in addition to frequency modulation [20]. Therefore, the zero-order beam is also amplitude modulated at the frequency $f_{FM}$ leading to interference detection. The addition of the modulator AOM#2 shifting the frequency of the probe beam ($f_2$ = 40 MHz) cancels this spurious signal by transposing the beat note frequency around 40 MHz Fig. 5(b). The beat note between the zero-orders of both AOM outputs is negligible.

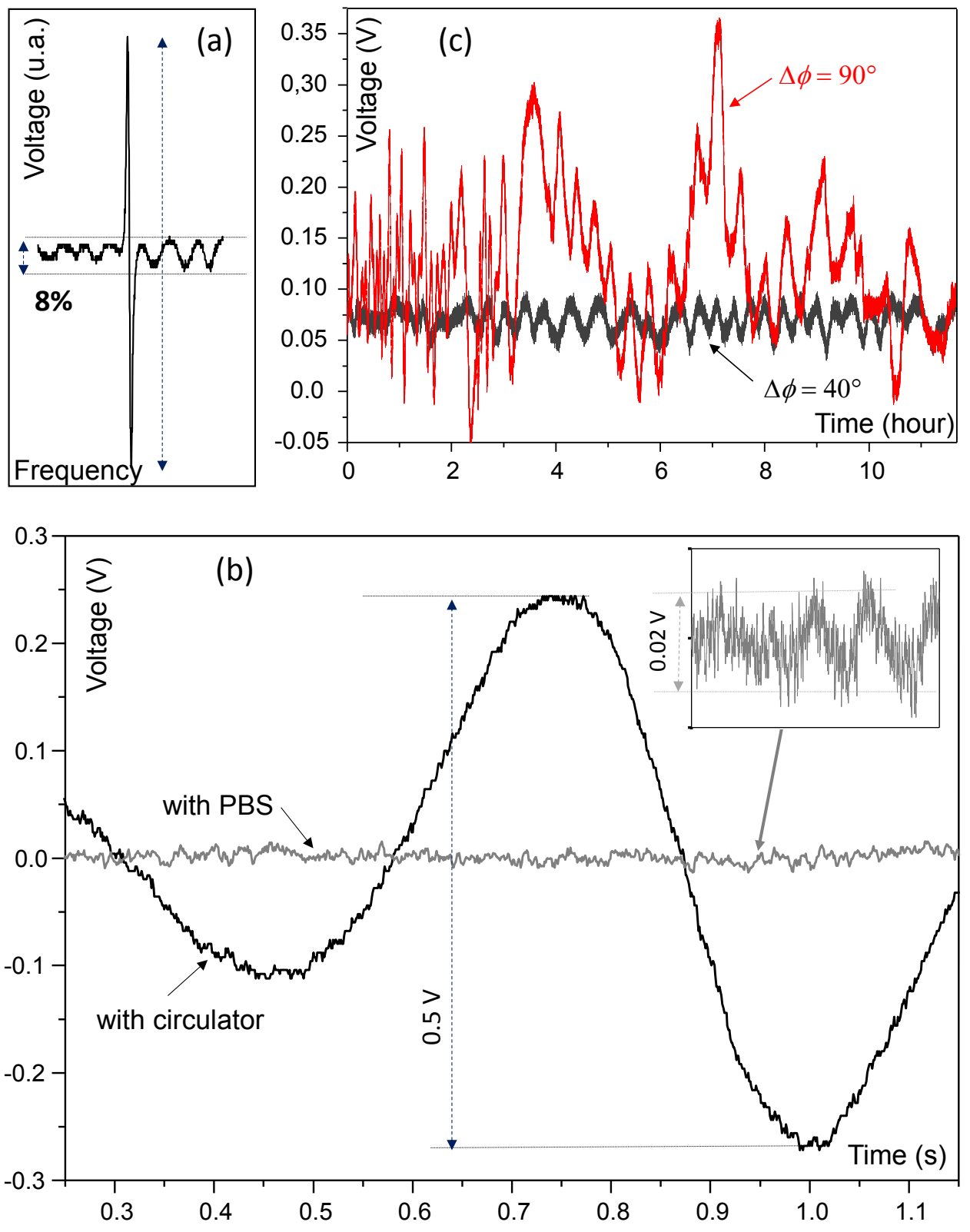

Fig. 4. Characterization of interference fringes in the setup: (a) Detected error signal without the frequency shift introduced by AOM#2; (b) Detected interference fringes with PBS#1 (grey curve) and with a circulator (black curve); (c) Detected interference fringes according to the detection phase (black curve: minimum amplitude - red curve: maximum amplitude).

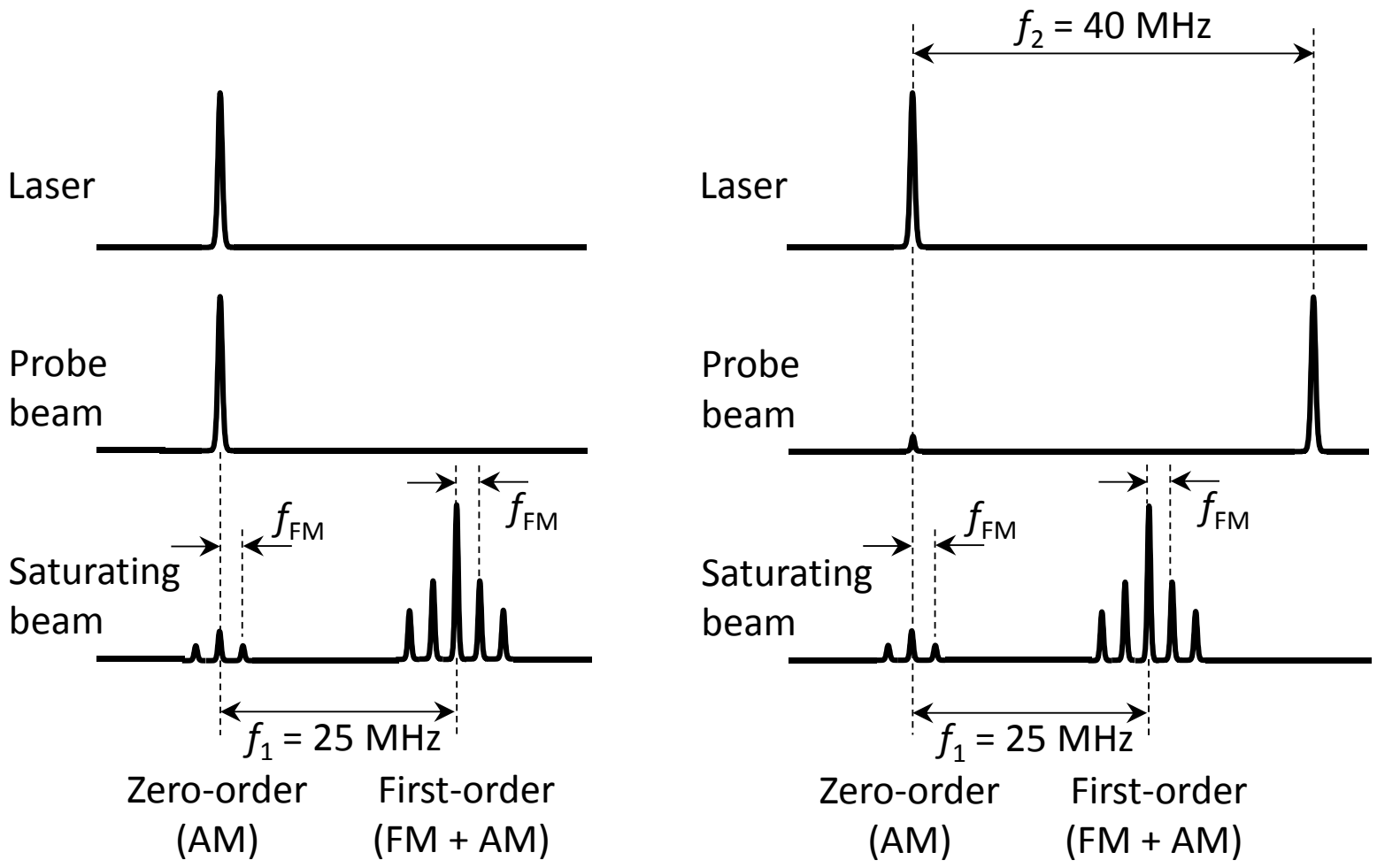

Fig. 5. Interference due to the zero-order of the AOM: (a) The probe beam is at the same frequency as the residual zero-order of the AOM that shifts and modulate the saturating beam frequency. The interference results from the detection at $f_{FM}$ of this zero-order that is amplitude modulated. (b) By shifting the frequency of the probe beam by 40 MHz, the beat note between the probe beam and the zero-order of the AOM is rejected at 40 MHz and therefore no interference is detected, the beat note between AOM zero-orders being negligible.

With AOM#2 in the setup, the residual interference fringes result from the direct detection of reflections from the saturating beam itself, namely the first-order output beam of AOM#1. It is frequency modulated at $f_{FM}$ = 200 kHz and carries a slight RAM as seen above. Fig. 4(b) shows the interference fringes detected when the probe beam is switch off (AOM#2 RF input OFF). The two traces correspond to the interference fringes recorded when PBS 1 is used to separate and route the optical beams, as in Fig. 1, and when PBS#1 is replaced by an optical circulator. In this latter case, a high level of interference is observed. The introduction of PBS#1 in the setup leads to an efficient rejection (by a factor of 25) of interference by filtering the saturating beam reflections according to its polarization. Nevertheless, it is seen that this rejection is limited by the polarization maintaining fiber (1m) at the PBS output presenting a polarization isolation of less than 30 dB between the two principal propagation axes.

Fig. 4(c) shows a ten-hour record of detected interference. The acquisition is performed with PBS#1, without the probe beam and when the saturating beam is replaced by the metrological reference at $\lambda$ref. Although the optical frequency of the source is perfectly constant, we observe a regular scrolling of the interference fringes during the recording. It causes the periodic shift of the zero crossing of the error signal for the laser frequency lock, leading to the bumps observed on the stability curves in Fig. 3. The detected fringes are strongly sensitive to the detection phase of the error signal. Their amplitude is minimized for a detection phase of 40° (black trace of Fig. 4(c)). The maximum interference amplitude is obtained for a detection phase of 90° (red curve in Fig. 4(c)) for which the detection of a slowly variable background is also observed.

We have seen above that the AOM output beam is both frequency and amplitude modulated (RAM). In addition, the RAM is strongly sensitive to the experimental environment. This leads to a variable background detection. Note that the phases associated with both modulations are different. For a detection phase that cancels the RAM, the interference fringes resulting only from the detection of the FM signal are obtained. This is the case of the black curve in Fig. 4(c) without background. For another detection phase, the interference fringes resulting from the RAM detection appear associated with a variable background, as for the red curve in Fig. 4(c).

This effect can be modeled simply by computing the detected signal resulting from the reflection by a low finesse cavity of a beam modulated simultaneously in frequency and amplitude. For the parameters of our experiments (FM index = 2, modulation frequency = 200 kHz, RAM = 2 $10^{-6}$), we found that the maximum amplitude of interference fringes due to RAM detection is 2.7 times greater than that due to the detection of the FM modulation alone. In addition, the interference resulting from RAM detection is superimposed to a continuous level while interference resulting from FM detection is centered at zero, which is consistent with the experimental results in Fig. 4(c).

The phase shift between FM and RAM generated by the AOM makes it possible to find a detection phase that cancels the signal due to RAM detection, and thus to minimize both interference and background while preserving an error signal of sufficient amplitude.

Consequently, RAM rejection and beam intensity control seem to be essential in such an experiment.

**5- Sensitivity to temperature**

The RAM generated by AOM#1 is temperature sensitive. This leads to a time-varying shift of the detection phase associated with the minimum of interference fringes (black curve in Fig. 4(c)) and thus a variation of the fringe offset. This is illustrated by Fig. 6 which shows the recording of the frequency fluctuations of the reference (Fig. 6(a)) and of the temperature in the thermal box (Fig. 6(b)) during a night with air conditioning in the lab corresponding to a regulation cycle of ~1800 s (blue curves) and without air conditioning (red curves). In both cases, the detection phase is initially adjusted to minimize the interference amplitude.

With air conditioning in the lab (blue curves), the temperature is sufficiently stable to maintain the initial adjustment leading to a constant average value of the frequency. The corresponding stability is given by the blue curve in Fig. 3. The periodic variation of the detected frequency (period ~2200 s) is reflected by the multiple bumps in Allan's variance as noted out above. Without particular drifts, we expect that for long integration times, the $\tau^{-1/2}$ trend reaches the stability level of $10^{-14}$.

When the air conditioning is turned off (red curves), the room temperature is less stable (variation of 3.5 °C during the night) and the frequency is observed to be unstable in the long term. A significant degradation of the stability of the frequency reference is observed (red curve in Fig. 3) leading to a trend in $\tau$ for integration time between 30 and 200 s and in $\tau^{1/2}$ beyond.

These measurements were performed with active beam intensity stabilization. The consequences of beam intensity variations are discussed in the next section.

**6- Sensitivity to beam power**

Another limitation of long-term stability of the reference is the sensitivity of the detected line center to the power of the probe beam and the saturating beam. We find a sensitivity of 65.7 Hz/mW for the probe beam and 40.7 Hz/mW for the saturating beam. The origin of instability of the intensity of the beams in the cell is the polarization fluctuations in fibers induced by the thermal variations. The PBS's convert them into intensity variations. This effect is compensated by controlling the intensity of both beams with VOA's. Residual fluctuations are much lower than 1 mW, which makes this effect negligible.

In order to quantify the sensitivity of our reference to beam power variations, we have measured the reference frequency with a stable temperature in the lab, but without correction of the power of the probe and saturating beams. The resulting stability is given by the black curve in Fig. 3. The period of interference fringes is modified as compared to the blue curve obtained with the beam power correction and the interference bump corresponding to the thermal sensitivity of the AOM is cancelled.

This result shows that the observed long-term instability is due to both beam power and RAM fluctuations, both due to thermal effects. In our setup, the control of the beam power and the external temperature rejects the long-term instabilities up to the limit given by the scrolling of interference fringes superimposed on the detected signal (blue curves of the Fig. 3).

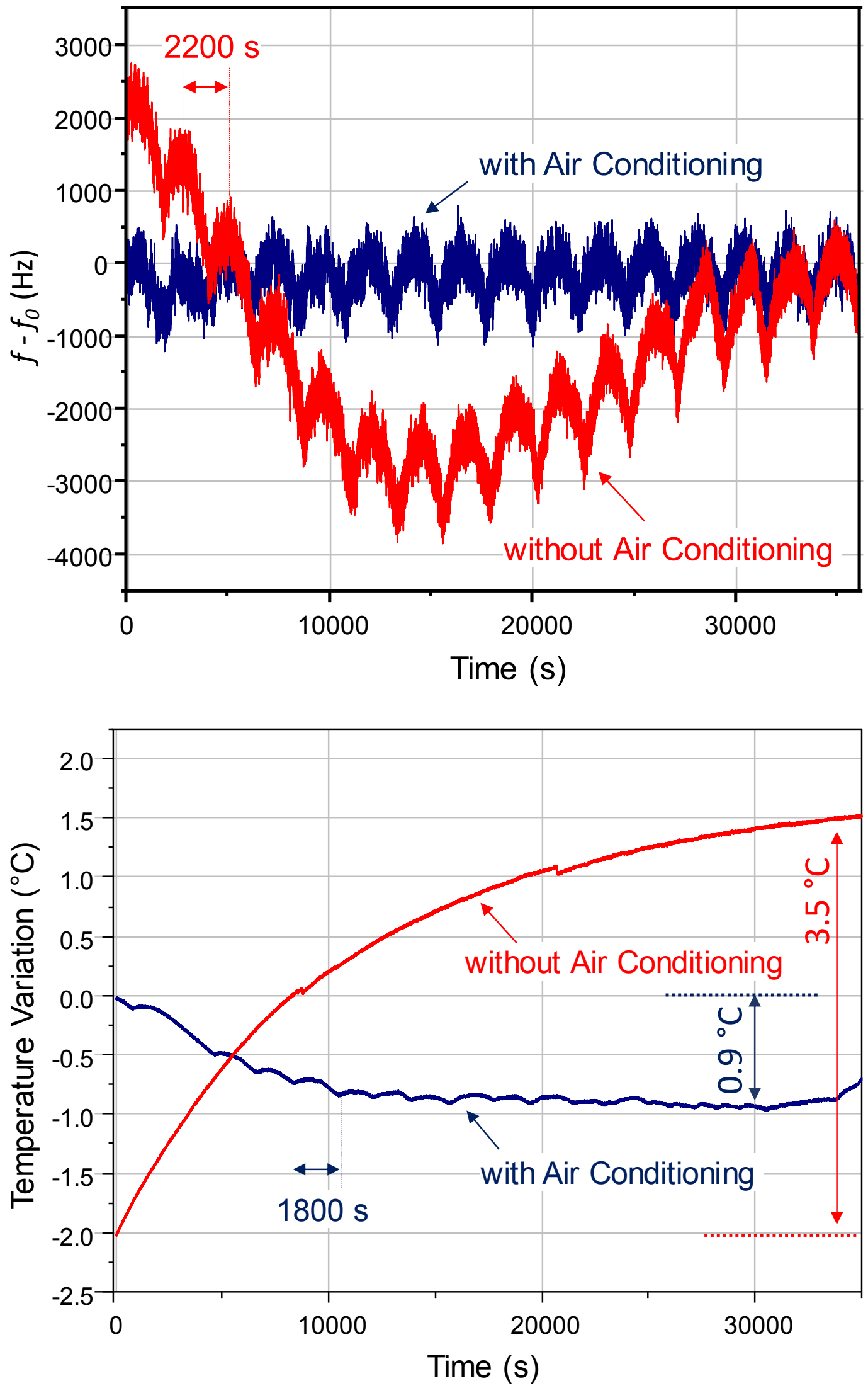

Fig. 6. Effect of temperature on the detected interference: (a) Frequency fluctuations of the beat-note between our reference and the REFIMEVE+ signal. The interference fringes scroll with a period of ~2200s; (b) Corresponding temperature in the thermal box with and without air conditioning system in the lab (cycle period of ~1800s).

## 7- Conclusion

We have described an optical frequency reference at 1.5 µm, based on standard fiber components. Frequency stabilization is achieved by locking the laser frequency to a saturated absorption signal in acetylene. The signal detection is performed in a cell by a modulation transfer method. Stability in the range of $10^{-13}$ between 1 and $10^4$ s is demonstrated.

We have investigated the origin of interference fringes and estimated their impact on the frequency stability of the reference. We have also highlighted the long-term effect of temperature and power fluctuations.

Setups based on fiber components have many advantages (compactness, lightness, robustness of optical alignments) which make them interesting, especially for the realization of transportable devices. However, even if some improvements could still be made to the setup (different powers of the pump and probe beams, replacement of connectors by splicing, choice of components), this work shows that interference constitutes an essential technological limitation of the fiber-based setups, limiting their stability in the $10^{-13}$ range. Moreover, we demonstrate that accurate temperature and power control may considerably limit RAM fluctuations and pave the way to the $10^{-14}$ range for long integration times.

**Acknowledgments**. We are grateful to H. Mouhamad and L. Malinge for their electronics support. This work was supported by the LABEX Cluster of Excellence FIRST-TF (ANR-10-LABX-48-01), within the Program "Investissements d'Avenir" operated by the French National Research Agency (ANR), and by PN-GRAM (Gravitation, Références, Astronomie, Métrologie). We acknowledge RENATER (Réseau National pour l'Enseignement et la Recherche), partner of the REFIMEVE+ project (ANR Equipex REFIMEVE+ ANR-11-EQPX-0039).